\documentclass[onecolumn,aps,showpacs,pra]{revtex4}
\usepackage{epsfig}
\usepackage[english]{babel}
\usepackage{latexsym}
\usepackage{graphics}
\usepackage{subfigure}
\usepackage{epsfig}
\usepackage{graphicx}
\usepackage{dcolumn}
\usepackage{amsmath}

\begin{document}
\title{{Modulations to molecular high order harmonic generation by electron de Broglie wave }
\footnotetext{$^{*}$ chen$\_$jing@iapcm.ac.cn;
  jingyun.fan@nist.gov}
 }
\author{J. Chen$^{1*}$, Y. J. Chen$^{1,3}$, J. Fan$^{2*}$, J. Liu$^{1}$, S. G. Chen$^1$, and X. T. He$^1$}

\date{\today}

\begin{abstract}
We present a new theory that the molecular high order harmonic generation
in an intense laser field is determined by molecular internal symmetry and
momentum distribution of the tunneling-ionized electron. The molecular
internal symmetry determines the quantum interference form of the returning
electron inside the molecule. The electron momentum distribution determines
the relative interference strength of each individual electron de Broglie
wave. All individual electron de Broglie wave interferences add together
to collectively modulate the molecular high harmonic generation.
We specifically discuss the suppression of the generation on adjacent
harmonic orders and the dependence of molecular high harmonic generation
on laser intensities and molecular axis alignment. Our theoretical results
are in good consistency with the experimental observations.

\end{abstract}
\affiliation{$^1$ Institute of Applied Physics and Computational Mathematics,
P.O. Box 8009 (28), Beijing 100088, P. R. China\\
$^2$ Optical Technology Division, National Institute of Standards
and Technology 100 Bureau Dr., Gaithersburg, MD 20899, USA\\
$^3$Graduate School, China Academy of Engineering Physics, P.O.
Box 8009-30, Beijing 100088, P. R. China\\}  \maketitle

Laser-molecule interaction has become an active research area in
high field physics. The molecular multi-center structure strongly
affects its interaction with the external laser field, resulting
in abundant new
physics\cite{tcc96,glng98,mbf00,nlhi03,aobw03,avtm04,lld03,dhl04,ilzn04,izls05,zsb05,lzm07,llx07}.
Recent studies show that the molecular orbital symmetry strongly
affects the single ionization\cite{tcc96,glng98,mbf00}; the nuclei
motion in molecular dissociation and Coulomb explosion can be
monitored by studying the electron rescattering
process\cite{nlhi03,aobw03}; the molecular orbital can be imaged
by studying the angular distribution of coincident ion pairs in
the molecular double ionization\cite{avtm04}; by studying the
spectral profile of molecular high order harmonic emission, one is
able to reconstruct the tomography of the molecular
structure\cite{ilzn04,izls05,lzm07,llx07}. Of particular interest
is the molecular high order harmonic generation (HHG). The
molecular multi-center structure provides the possibility to
actively modulate the molecular HHG to generate a single
attosecond pulse.

The ionization process of a molecule in an intense laser field is
described using the well-known simple-man's model\cite{c93}. First
one molecular electron tunnels through the potential barrier
assisted by the laser field, then the electron in the continuum
state may return to the molecular ion as the laser field reverses
its phase. Different from the rescattering process in the
laser-atom interaction, the multiple atomic sites inside the
molecular ion provide multiple options for the electron
recombination. The path-differences of the returning electron to
the multiple atomic sites result in the interference of electron
de Broglie wave ($e^{i{\bf k}\cdot{\bf r}}$) which significantly
affects the molecular ionization and HHG. This was first noticed
by Lein and his colleagues in studying the interaction of laser
with molecules\cite{lein02}. Lein \emph{et al}. showed that when
recaptured into the bonding orbit of a molecule, the electron
waves from the two atomic site emitters (with positions
$\textbf{r}_1$ and $\textbf{r}_2$) interfere as,
\begin{equation}
 e^{i\textbf{k}\cdot\textbf{r}_1}+e^{i\textbf{k}\cdot\textbf{r}_2},\label{a}
\end{equation}
which results in constructive or destructive interference when
${\bf k}\cdot{\bf R}$ (${\bf R}={\bf r}_1-{\bf r}_2$) is a
multiple or a multiple and a half of the laser wavelength. As a
result, the $n^{th}$-order molecular harmonic emission ($n\omega =
E_k$, $E_k=k^2/2$ ) is enhanced (or suppressed).

Recently two experiments were carried out independently to study
the HHG of $CO_2$ in an intense laser field with similar
experimental configurations only differing in laser intensities.
The suppression of HHG was observed for a range of adjacent
harmonic orders in both experiments\cite{kms05,vcb05}; while the
suppressed harmonic spectral ranges were displaced by about 8
harmonic orders with respect to each other. This cannot be
explained by Eq. (\ref{a}).

In this letter, we present a new theory based on the modified
molecular Lewenstein model to describe the laser-molecular
interaction. We show that similar to the Young's two-slit
interference, the returning electron (tunneling-ionized in the
external laser field) wave interferes on the multiple atomic sites
inside the molecule ion. The returning electron generally has a
momentum (${\bf k}$) distribution. Each single electron momentum
state (the electron de Broglie wave) self-interferes on the
multiple atomic sites. All interferences of different electron
momentum states (${\bf k}$) add together to affect the molecular
ionization and HHG. The interference structure is uniquely
determined by the molecular structure. The electron momentum
distribution is a function of laser intensity and molecular axis
alignment angle in the external laser field. Thus the molecular
HHG changes as laser intensity or molecular axis alignment varies.
We also show that molecular HHG at adjacent harmonic orders has
similar dependence on electron momentum distribution, thus the
suppression or enhancement on molecular HHG occurs over a broad
range of adjacent harmonic orders. Our theoretical results agree
well with experimental observations.

To directly compare with experimental observations, we study the
HHG of $CO_2$ which has anti-bonding orbital as the
highest-occupied state and can be approximately considered as a
two-center system when it interacts with the external laser
field\cite{kms05,vcb05}. Using the molecular Lewenstein model, the
dipole acceleration of $CO_2$ to yield the HHG radiation in an
external laser field (polarized along ${\bf x}$) is written
as\cite{ltl06,cl06,ccl06,ccl06pra},
\[
a_x(t)=i\int d^3{\bf k}|\Phi _i({\bf k})|^2{\int_0^t}dt^{\prime
}A(t^{\prime })
\]
\begin{equation}
\times E(t)(2\exp [-iS_0({\bf k},t,t^{\prime })]- \exp [-iS_1({\bf k}%
,t,t^{\prime })]- \exp [-iS_2({\bf k},t,t^{\prime
})])+c.c.,\label{c}
\end{equation}
where $A(t)$ and $E(t)$ are vector potential and electric field of
the laser pulse. Beside the action $S_0({\bf k},t,t^{\prime })
={\int_{t^{\prime }}^t}dt^{\prime \prime }(\frac 12 {\bf k}^2-{\bf
k}\cdot{\bf A}(t^{\prime \prime })+I_p)$, two additional action
terms $S_1({\bf k},t,t^{\prime }) =S_0({\bf k},t,t^{\prime })-{\bf
k}\cdot {\bf R}$  and $S_2({\bf k},t,t^{\prime }) =S_0({\bf
k},t,t^{\prime })+{\bf k}\cdot {\bf R}$  are introduced accounting
for the effects induced by the two O-atomic sites. The three phase
terms carrying these actions interfere with each other and
modulate the molecular HHG\cite{cl06,ccl06,ccl06pra,cc07}. $\Phi
_i({\bf k})$ is the Fourier transformation of the atomic
wavefunction $\varphi _i$ under \emph{linear combination of atomic
orbitals-molecular orbitals} (LCAO-MO) approximation in which the
molecular wavefunction is expressed as $\psi _i=\varphi _i({\bf
r},-{\bf R} /2)- \varphi _i({\bf r},{\bf R}/2)$.

After adopting pole approximation\cite{ccl06} and performing
time-integral in Eq. (\ref{c}), the Fourier transformation of Eq.
(\ref{c}) gives the HHG spectral amplitude of
$CO_2$\cite{ccl06pra,cc07}
\begin{equation}
 S(n)\propto\sum_{l,m}|(e^{i\textbf{k}^{\prime}\cdot\frac{\textbf{R}}{2}}-e^{-i\textbf{k}^{\prime}\cdot\frac{\textbf{R}}{2}})\Phi _i({\bf k}^{\prime})|^2
 J_{l}(\frac{-\mathbf{k}\cdot{\mathbf{A}_{0}}}{{\omega}})J_{m}(\frac{\mathbf{k}\cdot{\mathbf{A}_{0}}}{{\omega}}),\label{h}
\end{equation}
where $J_l()$ is the $l^{th}$ order Bessel function, $m$ and $l$
are numbers of photons that the electron absorbs or emits at
ionization or recombination, with $n=m\pm 1+l$. The electron's
momentum is now given as $k^2/2=(m\pm 1)\omega-I_p$, where $I_p$
is the ionization potential of the molecule. Considering the
acceleration effect by bound potential when the electron is in the
vicinity of the core, the modified electron momentum $k^{\prime}$
(parallel to $k$) is given as $k^{\prime
2}/2=k^2/2+I_p$\cite{cc07}.

Eq. (\ref{h}) shows that the molecular HHG is completely
determined by the structural factor(
$e^{i\textbf{k}^{\prime}\cdot\frac{\textbf{R}}{2}}
-e^{-i\textbf{k}^{\prime}\cdot\frac{\textbf{R}}{2}}$) which
corresponds to the molecular internal symmetry (here it describes
the anti-bonding orbital of $CO_2$) and the momentum distribution
[the rest part of Eq. (\ref{h})] of the rescattering electron
which is a function of the laser intensity and molecular axis
alignment angle (with respect to the direction of the external
laser field). Eq. (\ref{h}) also shows that the electron momentum
distribution and consequently the molecular HHG spectral profile
vary for different laser intensity and molecular alignment angle.
In the following, we numerically study the HHG of $CO_2$ under
various experimental conditions and compare the results with
available experimental data.

We first study the contribution to molecular HHG from electron
momentum states with momentum ${\bf k}$ parallel to ${\bf E}$,
because they were considered to contribute the most to the HHG. As
shown in Fig. 1, for both harmonic orders $S(n = 13)$ [Fig. 1(a)]
and $S(n = 23)$ [Fig. 1(b)], contributions from individual
electron momentum states (with ${\bf k}$  parallel to ${\bf E}$)
are significantly suppressed for smaller $k$ but remain unchanged
for big $k$, when considering the interference effect that is due
to the anti-bonding symmetry of $CO_2$. An integral of $\int
d\Omega\int^{k}_{0}k^2dk$ was then conducted to consider
contributions from all electron momentum states. Both $S(n = 13)$
[Fig. 1(c)] and $S(n = 23)$ [Fig. 1(d)] are constant for big $k$,
indicating that the contribution from big-$k$ electron momentum
states is negligible. When considering the interference effect,
$S(n = 13)$ is relatively suppressed while $S(n = 23)$ is
relatively enhanced.

Next we numerically study the variation of HHG spectral profile of
$CO_2$ as a function of laser intensity and molecular alignment
angle $\theta$. For fixed laser intensity, the HHG spectral
profiles vary for different values of  $\theta$ as shown in Figs.
2(a) and (b) in which the plots are generated with $I =
1.5\times10^{14}\ W/cm^2$ and $I = 2\times10^{14}\ W/cm^2$,
respectively. Eq. (\ref{h}) shows that harmonic emissions at
adjacent orders have similar dependence on electron momentum
contribution. Consequently similar modulations occur on adjacent
HHG orders. As shown in Fig. 2(c), with $I = 2\times10^{14}\
W/cm^2$ and $\theta = 30^{\circ}$, HHG is suppressed at a range of
adjacent harmonic orders centering at $n = 25$. The center shifts
to $n = 35$ for $I = 3\times10^{14}\ W/cm^2$ [Fig. 2(d)],
qualitatively agreeing with recent experimental
observations\cite{kms05,vcb05}. (Intensity stabilization and
calibration of intense ultrashort laser pulse remain to be a
challenge in the experiment.)

To clearly exhibit the dependence of molecular HHG on the laser
intensity, we numerically study the HHG at two harmonic orders.
For $n = 13$, our numerical results show that the harmonic
generation is suppressed at $I = 1.5\times10^{14}\ W/cm^2$
[Fig.3(a)] and enhanced at $I = 2\times10^{14}\ W/cm^2$
[Fig.3(b)]. For $n = 23$, we show the suppression of harmonic
generation  at $I = 2\times10^{14}\ W/cm^2$ [Fig.3(c)] and
enhancement at $I = 3\times10^{14}\ W/cm^2$ [Fig.3(d)].

Lastly, it is interesting to find out under what kind of condition
the single electron momentum picture given by Lein \emph{et al.}
may be used in predicting the suppression position in the HHG
spectrum. Examining Eq. (\ref{c}) shows that if  ${\bf k}\cdot{\bf
R}$ (describing the molecular structure) is much smaller than the
action $S_0$ (corresponding to the electron trajectory), then
$S_0$ can be extracted out as a common factor.

The Fourier-transformation of Eq. (\ref{c}) can now be simplified
into Eq. (\ref{a}). Assuming ${\bf A}(t)=\frac{{\bf
E}_0}{\omega}cos(\omega t)$, we have $S_0\sim \bar{E}_kT$, where
$T$ is the period of the laser field and $\bar{E}_k$is the average
kinetic energy of the electron with $\bar{E}_k\simeq
\frac{E_0^2}{2\cdot4\omega^2}$ and $k\simeq E_0/\omega$. Then we
reach such a relation
\begin{equation}
 R\ll\alpha=E_0/\omega^2,\label{condition}
\end{equation}
where $\alpha$ is the electron's quiver amplitude in the laser
field. Eq. (\ref{condition}) shows that if the electron's quiver
amplitude is much larger than the molecular internuclear distance,
or in other words, the time difference for the rescattering
electron to recombine with the different atomic sites are
negligible comparing to the time for electron being in the
continuum state, then Eq. (\ref{a}) can be used to approximately
describe the molecular HHG process. Comparing the internuclear
distances $R = 2.0\ a.u.$ for $H_2^+$ and $R = 4.4\ a.u.$ for
$CO_2$ with the electron's quiver amplitude of $\alpha= 21\ a.u.$
in a laser field with intensity of $I = 1.5\times10^{14}\ W/cm^2$,
it is easy to understand that Eq. (\ref{a}) gives a better
estimation to the suppression of HHG of $H_2^+$\cite{lein02} than
to the suppression of HHG of $CO_2$.

In conclusion, we develop a new theory about the molecular HHG in
an intense laser field. We illustrate that the molecular HHG at
each order is contributed by a group of interference effects on
the Young's slit (multiple-atomic sites) with each produced by an
independent electron de Broglie wave (eigen electron momentum
state) and weighed by the electron momentum state amplitude. We
show that the single electron momentum picture illustrated by Lein
\emph{et al}. is only an approximation of our theory at high laser
intensity where the electron's quiver motion amplitude is much
larger than the molecular internuclear distance. Our theoretical
analysis shows that the difference in the observed suppressed HHG
spectral ranges in recent experiments is attributed to the
different laser intensities applied in the measurement. We further
suggest that the molecular HHG in an intense laser field may be
generally described in the form of Eq. (\ref{h}) by simply
replacing the interference factor with the appropriate structure
factor of the molecule in use.

This work was supported by the National
Natural Science Foundation of China under Grant No. 10574019, 973
research program No. 2006CB806000, ICF research fund under Grant
No. 2004AA84ts08, CAEP Foundation No. 2006z0202 and the Multidisciplinary
University Research Initiative Center for Photonic Quantum Information Systems
(Army Research Office/DTO program DAAD19-03-1-0199).

\begin{figure}[tbh]
\begin{center}
\rotatebox{0}{\resizebox *{7.5cm}{5.cm} {\includegraphics
{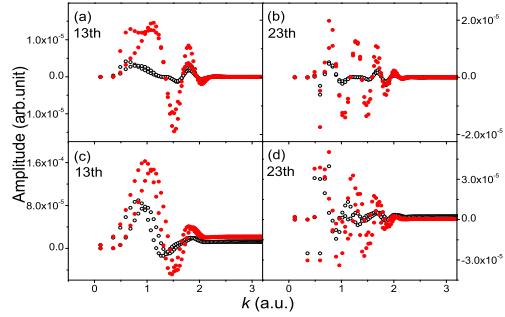}}}
\end{center}
\caption{ Calculated distributions of contributions to the HHG of
$CO_2$ at $n =13$ (a) and $n =23$ (b) from individual electron
momentum state with ${\bf k}$ parallel to ${\bf E}$, considering
(black open dots) and without considering (red filled dots)
interference effect. Integrated contributions from various
electron momentum states ($\int d\Omega\int^{k}_{0}k^2dk$) to the
HHG of $CO_2$ at $n =13$ (c) and $n =23$ (d), considering (black
open dots) and without considering (red filled dots) interference
effect. $I = 1.5\times10^{14}\ W/cm^2$ and $\lambda=800nm$. (The
double-line-structure is due to the double values of $k$ with
$k^2/2=(m\pm 1)\omega-I_p$.)} \label{fig.1}
\end{figure}

\begin{figure}[tbh]
\begin{center}
\rotatebox{0}{\resizebox *{7.5cm}{5.cm} {\includegraphics
{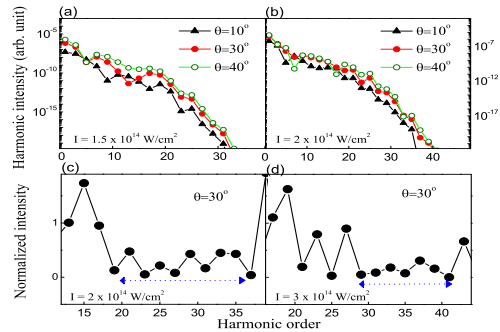}}}
\end{center}
\caption{ Calculated HHG power spectra of $CO_2$ at different
molecular alignment angles with (a) $I = 1.5\times10^{14}\ W/cm^2$
and (b) $I = 2\times10^{14}\ W/cm^2$. Calculated HHG suppression
(spectra normalized to that averaging over all angles) with (c) $I
= 2\times10^{14}\ W/cm^2$ and (d) $I = 3\times10^{14}\ W/cm^2$.
$\lambda=800nm$.} \label{fig.2}
\end{figure}

\begin{figure}[tbh]
\begin{center}
\rotatebox{0}{\resizebox *{7.5cm}{5.cm} {\includegraphics
{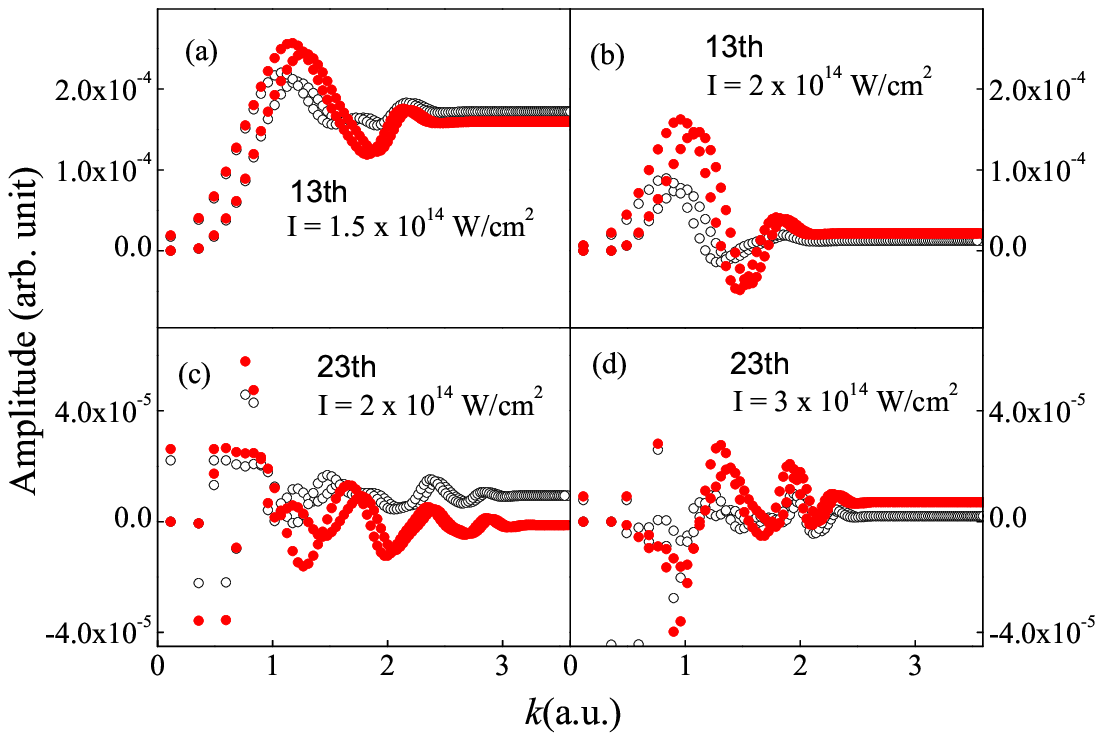}}}
\end{center}
\caption{
 Integrated contributions from various electron momentum states ($\int d\Omega\int^{k}_{0}k^2dk$)
 to the HHG of $CO_2$, considering (black open dots) and without considering
 (red filled dots) interference effect. (a) $n = 13$ with $I = 1.5\times10^{14}\ W/cm^2$
 and (b) $n = 13$ and $I = 2\times10^{14}\ W/cm^2$, (c) $n = 23$ with $I = 2\times10^{14}\ W/cm^2$
 and (d) $n = 23$ and $I = 3\times10^{14}\ W/cm^2$. $\lambda=800nm$.} \label{fig.3}
\end{figure}


\begin{thebibliography}{2}
\bibitem{tcc96}Talebpour A, Chien C Y and Chin S L 1996 J. Phys. B {\bf 29} L677
\bibitem{glng98} Guo C, Li M, Nibarger J P and Gibson G N 1998 Phys.
Rev. A {\bf 58} R4271
\bibitem{mbf00} Muth-B\"{o}hm J, Becker A
and Faisal F H M 2000  Phys. Rev. Lett. \textbf{85} 2280
\bibitem{nlhi03} Niikura H {\it et al} 2003 Nature \textbf{421}
826
\bibitem{aobw03} Alnaser A S {\it et al} 2003 Phys. Rev. Lett. \textbf{91}
163002
\bibitem{avtm04} Alnaser A S {\it et al} 2004 Phys. Rev. Lett. \textbf{93}
113003
\bibitem{lld03} Litvinyuk I V  {\it et al} 2003 Phys. Rev. Lett. \textbf{90}
233003
\bibitem{zsb05} Zeidler D {\it et al} 2005 Phys. Rev. Lett. \textbf{95}
203003
\bibitem{ilzn04} Itatani J {\it et al} 2004 Nature \textbf{432}
867
\bibitem{dhl04} de Nalda R {\it et al} 2004 Phys. Rev. A {\bf 69} 031804(R)
\bibitem{izls05}Itatani J {\it et al} 2005 Phys. Rev. Lett. {\bf 94} 123902
\bibitem{lzm07} Levesque J{\it et al} 2007 Phys. Rev. Lett. \textbf{98}
183903
\bibitem{llx07} Le Van-Hoang {\it et al} 2007 Phy. Rev. A \textbf{76} 013414
\bibitem{c93} Corkum P B 1993 Phys. Rev. Lett. {\bf 71} 1994
\bibitem{lein02} Lein M {\it et al} 2002 Phys. Rev. Lett. \textbf{88}
183903\\Lein M {\it et al} 2002 Phys. Rev. A \textbf{66} 023805\\
Lein M {\it et al} 2003 Phys. Rev. A \textbf{67} 023819
\bibitem{kms05} Kanai T, Minemoto S and Sakai H 2005 Nature \textbf{435}
470-473
\bibitem{vcb05} Vozzi C {\it et al} 2005 Phys. Rev. Lett. \textbf{95}
153902\\ Vozzi C {\it et al} 2006 J. Phys. B \textbf{39} S457
\bibitem{ltl06}Le Anh-Thu, Tong X M  and Lin C D 2006
Phy. Rev. A \textbf{73} 041402(R)
\bibitem{cl06}
Chirila C C and Lein M 2006 Phys. Rev. A {\bf 73} 023410
\bibitem{ccl06}
Chen J, Chu Shih-I  and Liu J 2006 J. Phys. B. \textbf{39}
4747-4758
\bibitem{ccl06pra} Chen Y J, Chen J and Liu J 2006 Phys. Rev. A. \textbf{74} 063405
\bibitem{cc07} Chen Y J {\it et al} Model analysis of two-center interference on high
harmonic generation, unpublished.




\end{thebibliography}
\end{document}